\documentclass[useAMS,usenatbib]{mn2e}

\usepackage{amssymb}
\usepackage{graphicx}
\usepackage{mathptmx}
\usepackage{amsmath}
\usepackage{color}

\title[Star Cluster Early Evolutionary Rates]{The mass function and dynamical mass of young star clusters: \\ Why their initial crossing-time matters crucially}

\author[G. Parmentier \& H. Baumgardt]
{Genevi\`eve Parmentier$^{1,2}$\thanks{E-mail: gparm@ari.uni-heidelberg.de - Present address: Astronomisches Rechen-Institut, Heidelberg University, M\"onchhofstr. 12-14, D-69120 Heidelberg, Germany} \& Holger Baumgardt$^{3}$ \\
$^{1}$ Max-Planck-Institut f\"ur Radioastronomie, Auf dem H\"ugel 69, D-53121 Bonn, Germany \\
$^{2}$ Argelander-Institut f\"ur Astronomie, Bonn Universit\"at, Auf dem H\"ugel 71, D-53121 Bonn, Germany \\
$^{3}$ School of Mathematics and Physics, The University of Queensland, Brisbane, QLD 4072, Australia
}

\begin{document}

\date{Accepted 2012 August 29.  Received 2012 July 16; in original form 2011 September 28
}
\pagerange{\pageref{firstpage}--\pageref{lastpage}} \pubyear{201?}

\maketitle

\label{firstpage}

\begin{abstract}
We highlight the impact of cluster-mass-dependent evolutionary rates upon the evolution of the cluster mass function during violent relaxation, that is, while clusters dynamically respond to the expulsion of their residual star-forming gas.  Mass-dependent evolutionary rates arise when the mean volume density of cluster-forming regions is mass-dependent.  In that case, even if the initial conditions are such that the cluster mass function at the end of violent relaxation has the same shape as the embedded-cluster mass function (i.e. infant weight-loss is mass-independent), the shape of the cluster mass function does change transiently {\it during} violent relaxation.  In contrast, for cluster-forming regions of constant mean volume density, the cluster mass function shape is preserved all through violent relaxation since all clusters then evolve at the same mass-independent rate.  
 
On the scale of individual clusters, we model the evolution of the ratio between the dynamical mass and luminous mass of a cluster after gas expulsion.  Specifically, we map the radial dependence of the time-scale for a star cluster to return to equilibrium.  We stress that fields-of-view a few pc in size only, typical of compact clusters with rapid evolutionary rates, are likely to reveal cluster regions which have returned to equilibrium even if the cluster experienced a major gas expulsion episode a few Myr earlier.  We provide models with the aperture and time expressed in units of the initial half-mass radius and initial crossing-time, respectively, so that our results can be applied to clusters with initial densities, sizes, and apertures different from ours.
\end{abstract}


\section{Introduction}
\label{sec:intro}
Star cluster formation conditions have a profound impact on the properties of young star cluster systems.  In particular, after massive star activity has cleansed a cluster-forming region (CFRg) of its residual star-forming gas, the mass fraction of stars bound to a cluster decreases steadily with time (infant weight-loss) until the cluster either reaches a new equilibrium, or is completely disrupted.  This dynamical response of a cluster to gas expulsion is called violent relaxation.  \citet{bau07} express the bound mass fraction of stars, $F_b$, as a  function of the CFRg star formation efficiency at the onset of gas expulsion, $SFE$, of the gas expulsion time-scale, $\tau_{gexp}/\tau_{cross}$, and of the impact of an external tidal field which they quantify through the ratio of the initial half-mass radius to the initial tidal radius of the embedded-cluster, $r_h/r_t$.  That is:
\begin{equation}
F_b = F_b (t/\tau_{cross},SFE,\tau_{gexp}/\tau_{cross}, r_h/r_t )\,,
\label{eq:fb}
\end{equation}      
where $F_b$ is defined as the mass fraction of stars within the cluster tidal radius at time $t$.  
Note that the time-scale for gas expulsion, $\tau_{gexp}$, and the time since gas expulsion, $t$, are both expressed in units of the CFRg crossing-time, $\tau_{cross}$.  This immediately implies that denser CFRgs give rise to faster-evolving clusters since $\tau_{cross} \propto \rho_{CFRg}^{-1/2}$, with $\rho_{CFRg}$ the CFRg mean volume density.  This property is the crux of the present contribution.  The higher the $SFE$, the longer the gas-expulsion time-scale $\tau_{gexp}/\tau_{cross}$, the weaker the tidal field impact $r_h/r_t$ (i.e. the deeper the cluster lies within its limiting tidal radius), the smaller the cluster infant-weight loss, $1-F_b$, at time $t/\tau_{cross}$.  Cluster violent relaxation therefore leaves an imprint on the age distribution \citep{par09a} and mass function of young star clusters \citep{kro02, bau08, par08b, par11}. \\

During the few 10s of Myr after gas expulsion, infant weight-loss drives the cluster mass function towards lower cluster masses.  The amplitude of this leftward shift as a function of time, and mass (i.e. does the cluster mass function shape vary with time?), constitutes a powerful probe into cluster formation conditions.  Most observational evidence gathered so far seem to support the scenario of mass-independent infant weight-loss, $1-F_{b}$, at that young age.    When plotted as the number of clusters per linear mass interval, the young cluster mass function in the present-day Universe is a power-law of slope $\simeq -2$, ${\rm d}N_{cl} \propto m_{cl}^{-2} {\rm d}m_{cl}$, irrespective of the cluster age \citep[e.g.][]{ll03,oey04,dow08,cha10}.  
It should be kept in mind that, in assessing the slope of power-law cluster mass functions, systematic errors (e.g. the errors inherent to the derivations of cluster mass estimates by different research groups) are significantly larger than the statistical ones (i.e. the formal fitting errors).  For instance, \citet{bau1X} find that the mass function slope of the Large Magellanic Cloud clusters younger than 200\,Myr is $-2.32\pm0.11$.  This is steeper than the slope reported by \citet{deg06} for a similar cluster age range, i.e. $-1.94\pm0.10$.  This simple example shows that to detect an intrinsic change of $\simeq 0.2$ in the cluster mass function slope is not  straightforward.
We note that through a careful analysis of the combined star cluster systems of several galaxies -- an approach which aims at decreasing statistical uncertainties at the high mass end -- \citet{lar09} shows that the cluster mass function obeys a Schechter function, rather than a featureless power-law, with a cut-off mass higher in starbursts and mergers than in quiescent spiral discs.  

The amount of mass-independent dissolution affecting young clusters remains heavily debated.  Based on the relation between the mass of the most massive cluster and the age range sampled, \citet{gie08} infer almost negligeable mass-independent dissolution, i.e. 20\,\% per age dex as compared to 90\,\% per age dex for \citet{cha10}.  We stress that these observational uncertainties do not affect the general conclusions we present here.   

\indent If infant weight-loss is mass-independent, this straightforwardly implies mass-independent $SFE$ and mass-independent $\tau_{gexp}/\tau_{cross}$, although the constraint on $\tau_{gexp}/\tau_{cross}$ is looser than for the $SFE$ \citep[see section 4.1 in][for a discussion]{par11}.  For a given external tidal field (e.g. a limited region of a given galaxy), it also implies the mass-independence of $r_h/r_t$.  This constraint is robustly satisfied if the mass-radius relation of CFRgs is one of constant mean volume density \citep{par11}, i.e. $r_{CFRg} \propto m_{CFRg}^{1/3}$, with $r_{CFRg}$ and $m_{CFRg}$ the radius and mass of CFRgs, respectively.\footnote{Note that molecular clumps show density gradients \citep{mue02, beu02}.  It is thus important to keep in mind that this criterion refers to the {\it mean} volume density of  CFRgs.  It does {\it not} imply that the volume density within a CFRg is uniform.}
Mass-independent $SFE$, $\tau_{gexp}/\tau_{cross}$ and $r_h/r_t$ therefore imply that the cluster mass functions at the onset and end of violent relaxation have the same shape, albeit different amplitudes due to cluster infant weight-loss.

A fourth, and so far overlooked, aspect is how the {\it cluster evolutionary rate} shapes the cluster mass function {\it during} violent relaxation.  If the mass-radius relation of CFRgs is one of constant mean surface density (i.e. $r_{CFRg} \propto m_{CFRg}^{1/2}$), more massive CFRgs have a lower mean volume density and, therefore, evolve on a slower time-scale.  Conversely, if CFRgs all have the same radius, $r_{CFRg}$, irrespective of mass, higher-mass CFRgs give rise to faster-evolving star clusters, as illustrated by fig.~1 of \citet{par10}.  In that example, the constant radius also implies that more massive CFRgs have deeper gravitational potential wells hence slower gas-expulsion time-scales.  This allows high-mass clusters to survive despite a low $SFE$ of $0.20$, while low-mass ones are disrupted.  The mass-dependent gas-expulsion time-scale thereby carves a turnover in the initially power-law cluster mass function.  The same figure also shows that, as clusters evolve after gas expulsion, transient patterns sets in the cluster mass function due to the mass-dependent evolutionary rate.  At an intermediate age of 15\,Myr, clusters less massive than $3000\,M_{\odot}$ (i.e. low-density clusters) have hardly evolved and their mass function sticks to the initial one.  That is, none of their stars have yet crossed their tidal boundary.  Conversely, clusters more massive than $10^4\,M_{\odot}$ (i.e. high density clusters) have completed their violent relaxation, and the bell-shaped cluster mass function already shows up in the high-mass regime.

In this contribution, we expand the work of \citet{par10}.  We highlight how the CFRg mass-radius relation affects the evolutionary rate of star clusters in dependence of their mass and, therefore, the cluster mass function evolution.  In particular, we quantify transient mass-dependent effects which do take place {\it even if the cluster mass function shapes at the onset and at the end of violent relaxation are alike.}  Although we illustrate our point with a power-law cluster mass function experiencing 90\,\% infant weight-loss, we emphasize that the cluster mass function patterns highlighted below do take place regardless of the initial cluster mass function shape and regardless of the infant weight-loss rate.  

While the cluster mass function evolution constitutes a good tracer of violent relaxation on the scale of star cluster {\it systems}, the dynamical evolution of an {\it individual} cluster after gas expulsion can be traced by the ratio of its dynamical-to-luminous mass estimates, $M_{dyn}/M_{lum}$.  Again, the shorter the initial crossing-time, the faster the return of the cluster to virial equilibrium where $M_{dyn}/M_{lum} \simeq 1$.  The question of whether a young cluster can be observed in virial equilibrium \citep[e.g. Wd~1, ][]{men07} despite having been put out of equilibrium by a significant gas-expulsion episode in the recent past is thus directly related to its initial crossing-time hence to the CFRg mass-radius relation.  As we shall see in the present work, the problem is complexified by the size of the aperture with which the cluster is observed.  For a given time-span after gas expulsion, smaller apertures are more likely to miss the fast-moving unbound stars, thereby decreasing the observed $M_{dyn}/M_{lum}$ ratios with respect to larger apertures.
 
The outline of the paper is as follows.  In Section \ref{sec:mod}, we quantify the interplay between the cluster mass function shape and the CFRg mass-radius relation in the case of a weak tidal field.  Section \ref{sec:ap} models the time-evolution of the cluster dynamical-to-luminous mass ratio, $M_{dyn}/M_{lum}$.  We demonstrate that the time-scale for a cluster to recover virial equilibrium after gas expulsion depends on the size of the aperture with which it is observed.  Our discussion helps understand why a few massive clusters are observed in virial equilibrium despite a young age of several Myr only.  
We present our conclusions in Section \ref{sec:conclu}.

\section{Transient Cluster Mass Functions}
\label{sec:mod}
The bound mass of a star cluster after gas expulsion obeys:
\begin{equation}
m_{cl}=F_b \times SFE \times m_{CFRg}\;.
\label{eq:mcl}
\end{equation}
We obtain $F_b$ by linearly interpolating the $N$-body model grid of \citet{bau07}
\footnote{The library of $N$-body models on which we build is described in detail in \citet{bau07}.  
It can be downloaded from: http://www.astro.uni-bonn.de/download/data/.  
The library consists of three sets of files: the `esc'-extension files describe the time evolution of the star bound fraction $F_b$, the `rad\_bound'-extension files quantify the time evolution of the three-dimensional Lagrangian radii of the bound stars, and the `rad\_bound\_2d'-extension files describe the time evolution of the projected Lagrangian radii of all cluster stars, i.e. bound and unbound unlike.}.  
This provides $F_b$ as a function of $t/\tau_{cross}$, $SFE$, $\tau_{gexp}/\tau_{cross}$ and $r_h/r_t$ (i.e. as in Eq.~\ref{eq:fb}).  To recover $F_b$, we have to express time in unit of $\tau_{cross}$.  For CFRgs with a volume density profile of slope $-1.7$ \citep{mue02,par11c}, we find that the crossing-time at the limiting radius $r_{CFRg}$ obeys\footnote
{Formally, the crossing-time relevant to the $N$-body time-unit is the crossing-time at the virial radius of the CFRg \citep{heg03}.  It is shorter than the crossing-time at the CFRg limiting radius given by Eq.~\ref{eq:tcr} by about one third only and we thus neglect that difference.  All through Section \ref{sec:mod}, the crossing-time of a CFRg is as defined by Eq.~\ref{eq:tcr}
}:  
\begin{equation}
\tau_{cross}[Myr] = 33 \left(\frac{(r_{CFRg} [pc])^{3}}{(m_{CFRg} [M_{\odot}])}\right)^{1/2}\;.
\label{eq:tcr}
\end{equation}

We consider a power-law CFRg mass function of slope $-2$, ${\rm d}N_{CFRg} \propto m_{CFRg}^{-2} {\rm d}m_{CFRg}$, totalizing $10^8\,M_{\odot}$ of dense star-forming gas.  Our aim is to highlight the impact of the sole evolutionary rate on the cluster mass function, that is, the impact of $t/\tau_{cross}$ in Eq.~\ref{eq:fb}.  To that purpose, we assume all other parameters ($SFE$, $\tau_{gexp}/\tau_{cross}$, $r_h/r_t$) to be mass-independent.  We take $SFE=0.35$ and $\tau_{gexp}/\tau_{cross}=0.01$ (explosive gas expulsion).  Note that a longer gas-expulsion time-scale leaves $F_b$ unaffected as long as $\tau_{gexp}/\tau_{cross} \lesssim 0.3$ \citep[see fig.~1 in][]{par08b}.

\begin{figure}
\includegraphics[width=\linewidth]{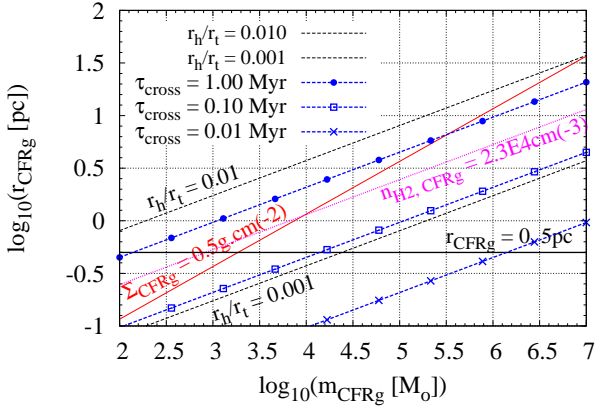}
\caption{Mass-radius diagram of cluster-forming regions (CFRg) with lines of constant mean number density ($n_{\rm H_2, CFRg}=2.3 \times 10^4\,cm^{-3}$, pink dotted line), constant mean surface density ($\Sigma_{CFRg}=0.5\,g.cm^{-2}$, red solid line), constant radius ($r_{CFRg}=0.5\,pc$, black solid horizontal line), and constant crossing-times (blue dashed lines with symbols; see key).  Illustrated are the mass-dependences of CFRg crossing-times in relation with their mass-radius relations, hence cluster-mass-dependent evolutionary rates.  \label{fig:mrr} }
\end{figure}

\begin{table}
\begin{center}
\caption{Adopted mass-radius relations for cluster-forming regions (CFRg).  Also given are the mean surface density, mean volume density, and radius of the $\Sigma _{CFRg}$, $\rho _{CFRg}$ and $r_{CFRg}$ models, respectively, as well as the corresponding CFRg crossing-times, $\tau_{cross}$, in dependence of $m_{CFRg}$.  When not quoted explicitly, units are pc, $M_{\odot}$ and Myr.  \label{tbl:mrr}}
\begin{tabular}{cccc}
\hline\hline
                      & $\Sigma _{CFRg}$      &  $\rho _{CFRg}$                   &  $r_{CFRg}$  \\ \hline
                      & 0.5g.cm$^{-2}$        &  $1.1\,10^{-19}$g.cm$^{-3}$       &  0.5pc       \\
$r_{CFRg}[pc]=$       & $0.01m_{CFRg}^{1/2}$  &  $0.05m_{CFRg}^{1/3}$             &  $0.5$ \\
$\tau_{cross}[Myr]=$  & $0.04m_{CFRg}^{1/4}$  &  $0.41$                           &  $11.7m_{CFRg}^{-1/2}$\\ \hline
\end{tabular}
\end{center}
\end{table}

In essence, the tidal field impact $r_h/r_t$ is mass-dependent if the CFRg mass-radius relation is not one of constant mean volume density \citep[eqs~8 and 10 in][]{par11}.  When $r_{CFRg} \propto m_{CFRg}^{1/2}$ (constant mean surface density), $r_h/r_t \propto m_{CFRg}^{1/6}$ and higher-mass clusters experience greater infant weight-loss through their tidal boundary.  In case of constant radius $r_{CFRg}$, $r_h/r_t \propto m_{CFRg}^{-1/3}$ and higher-mass clusters are more resilient to the external tidal field.  To `switch off' those mass-dependent effects, we assume that the environment exerts upon star clusters a tidal field weak enough so that $r_h/r_t \leq 0.01$ over the entire mass range.  When $r_h/r_t \leq 0.01$, clusters respond to gas expulsion as if in a tidal-field-free environment \citep{bau07}.  In other words, the bound fraction $F_b$ does not respond to variations in $r_h/r_t$ as long as $r_h/r_t \leq 0.01$.  

Figure~\ref{fig:mrr} illustrates the CFRg mass-radius relations encompassed by our simulations.  We refer to them as the $\Sigma_{CFRg}$ (constant mean surface density), $\rho_{CFRg}$ (constant mean volume density) and $r_{CFRg}$ (constant radius) models.  They all lie below the line $r_h/r_t= 0.01$.  As just explained, this cancels any tidally-induced mass-dependent effect upon $F_b$ despite $\Sigma_{CFRg}$ and $r_{CFRg}$ models having mass-varying $r_h/r_t$.  Figure~\ref{fig:mrr} also shows lines of iso-$\tau_{cross}$ (see Eq.~\ref{eq:tcr}).  By virtue of their mass-dependent volume densities, the $r_{CFRg}$ and $\Sigma_{CFRg}$ models give rise to crossing-times shorter and longer, respectively, towards higher masses.  Model properties, along with their corresponding crossing-time in dependence of mass, are given in Table~\ref{tbl:mrr}.
The adopted mass-radius relations are representative of star-forming molecular clumps \citep[see middle and bottom panels of fig.~1 in][]{par11}.  We stress, however, that molecular clump sizes are often FWHM measurements which, therefore, do not necessarily represent the outer boundary of the star clusters formed by these clumps (see \citet{par11b} and \citet{par11c} for applications of this concept).  Other normalizations of the mass-radius relations would simply affect $\tau_{cross}$ hence the age -- but not the shapes -- of the cluster mass functions we obtain below.

The stellar mass of an embedded cluster obeys: $m_{ecl}=SFE \times m_{CFRg}$.  With mass-independent $SFE$, the embedded-cluster mass function is simply the CFRg mass function shifted horizontally by $log_{10}(SFE) \simeq -0.5$.  In addition, with mass-independent $\tau_{gexp}/\tau_{cross}$, and $r_h/r_t \lesssim 0.01$, the cluster mass function at the end of violent relaxation is the embedded-cluster mass function shifted horizontally by $log_{10}(F_{bound})$, where $F_{bound}$ is the mass-independent bound mass fraction at the end of violent relaxation. 
We refer to the bound fraction at the {\it end} of violent relaxation as $F_{bound}$ to make it distinct from $F_b$, the bound fraction at {\it any time} during violent relaxation.

\begin{figure}
\includegraphics[width=\linewidth]{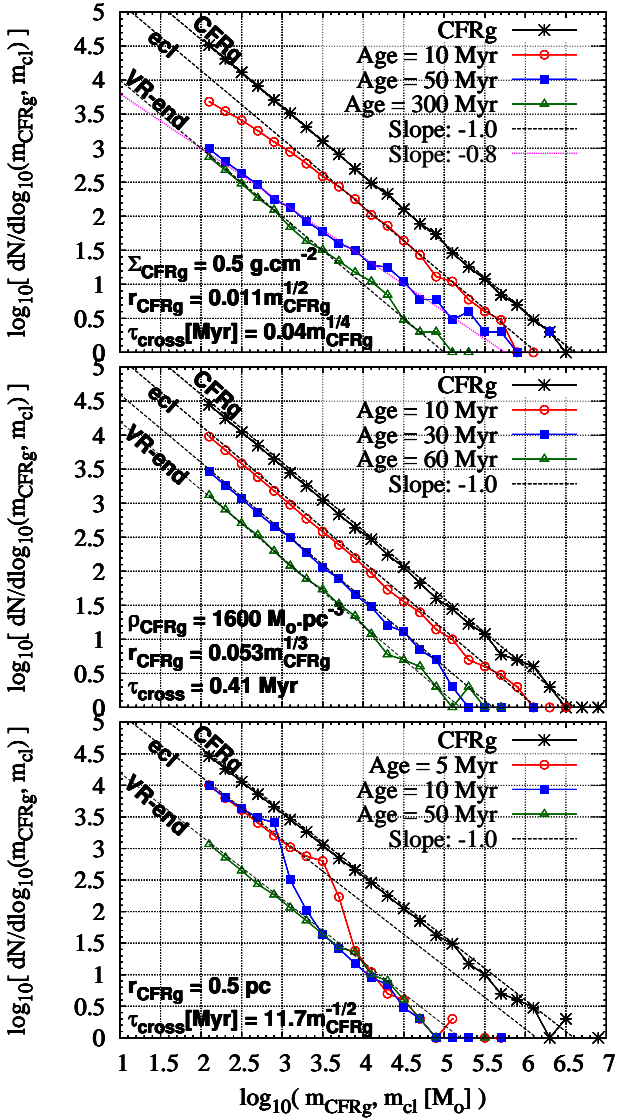}
\caption{Evolution of the cluster mass function through violent relaxation for three CFRg mass-radius relations: constant mean surface density ($\Sigma_{CFRg}$), constant mean volume density ($\rho_{CFRg}$) and constant radius ($r_{CFRg}$), from top to bottom, respectively.  Each panel depicts the CFRg mass function (solid black line with asterisks) and cluster mass functions at three distinct ages (see key).  The mass of a cluster is defined as the stellar mass enclosed within its tidal radius at the age of relevance.  In top and bottom panels, violent relaxation transiently distorts the cluster mass function compared to the CFRg mass function because of cluster-mass-dependent evolutionary rates. \label{fig:MF} }
\end{figure}

For the model parameters adopted here -- $SFE = 0.35$, $\tau_{gexp}/\tau_{cross} = 0.01$, $r_h/r_t \lesssim 0.01$ -- $F_{bound}=0.11$.  That is, the power-law mass functions of CFRgs, embedded clusters and post-violent-relaxation clusters all have the same slope.  In Fig.~\ref{fig:MF}, they are schematically depicted as the (black) dashed lines labelled `CFRg', `ecl' and `VR-end', respectively.  Their slope is $-1$ since Fig.~\ref{fig:MF} represents the number of clusters per constant logarithmic mass bin, i.e. $dN/d\log m$.  In all this section, the slope of a cluster mass function refers to the slope of $dN/d\log m$ (e.g. ${\rm d}N_{cl} \propto m_{cl}^{-2} {\rm d}m_{cl}$ has a slope of $-1$).

Over the course of violent relaxation, however, the cluster mass function shape is modified if the evolutionary rate is mass-dependent.  For the $\Sigma_{CFRg}$ model, clusters of higher mass evolve more slowly than low-mass ones, which renders the cluster mass function transiently shallower than the embedded-cluster mass function.  Conversely, for the $r_{CFRg}$ model, more massive clusters evolve faster and we thus expect the opposite behaviour.

The main results of our simulations are presented in Fig.~\ref{fig:MF}, with the $\Sigma _{CFRg}$, $\rho _{CFRg}$ and $r_{CFRg}$ models in top, middle and bottom panels, respectively.  Each panel presents the `parent' CFRg mass function (solid (black) line with  asteriks) and model cluster mass functions at three distinct ages (see panel key).  To limit the Poisson noise of our Monte-Carlo simulations, each displayed mass function is the median of a set of 100 simulations.  In the top two panels, the youngest cluster mass function (red solid line with open circles) does not differ from the embedded-cluster mass function.  That is, clusters are still dynamically too young for the first stars due to be unbound to have reached their limiting tidal radius.  
 
\begin{figure}
\includegraphics[width=\linewidth]{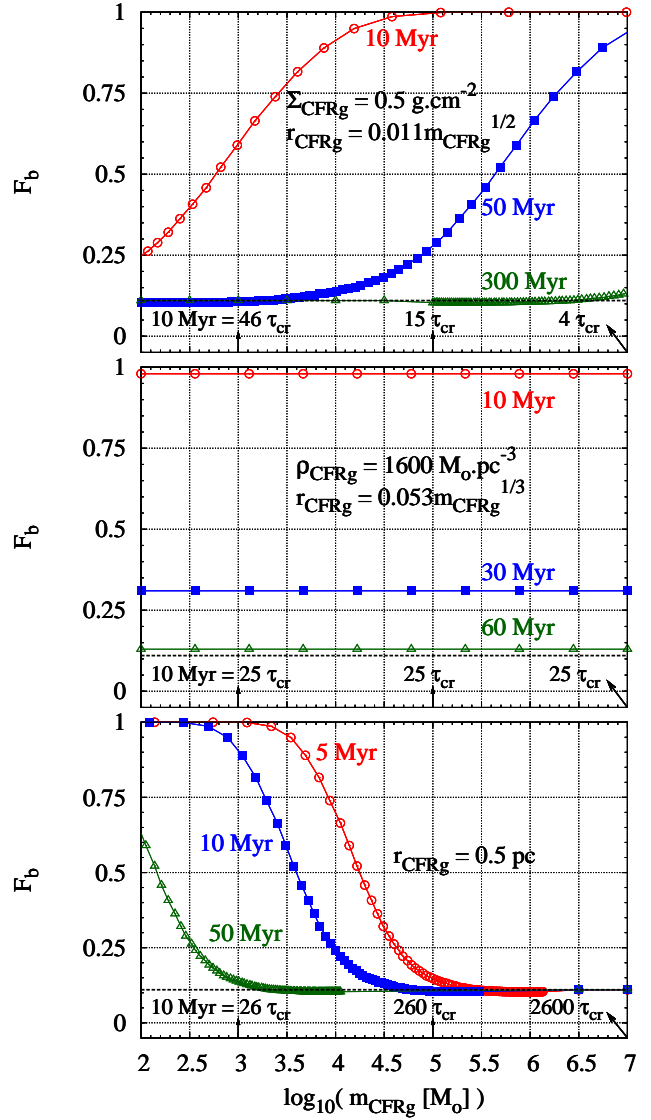}
\caption{Bound fraction $F_b$ as a function of CFRg mass and cluster age.  CFRg mass-radius relations and cluster ages are identical to Fig.~\ref{fig:MF} (same colour/symbol codings).  
For the $\Sigma_{CFRg}$ model (top panel), massive clusters evolve more slowly than low-mass ones, while the opposite is true for constant radius (bottom panel).  For CFRgs of constant mean volume density (middle panel), the cluster evolutionary rate is mass-independent
\label{fig:mfb} }
\end{figure}

As expected, for the $\rho_{CFRg}$ model (middle panel), the shape of the cluster mass function does not change with time since $\tau_{cross}$ is mass-independent.  For the mean volume density $\rho_{CFRg}$ assumed here, violent relaxation is over by an age of 60\,Myr (see also Fig.~\ref{fig:mfb}), that is, all stars due to be unbound have crossed the cluster instantaneous tidal radius\footnote{It is worth keeping in mind that the cluster tidal radius itself is shrinking with time since it depends on the {\it bound} cluster mass, i.e. $r_t \propto m_{cl}^{1/3}$.  Hence the terminology `instantaneous tidal radius' adopted here}.  For the CFRg surface density considered in the top panel, violent relaxation is completed by clusters of all masses by an age of 300\,Myr.
While violent relaxation proceeds, the cluster mass function becomes shallower (slope $\simeq -0.8$; solid blue line with plain squares) than the CFRg mass function.  The opposite behaviour characterizes the $r_{CFRg}$ model (bottom panel), with more massive clusters reaching their post-violent-relaxation  mass at younger ages than low-mass ones.  In this case, the transient shape of the cluster mass function can be described as a 3-segment sequence.  Note that the cluster mass functions at the end of violent relaxation (green lines with open triangles) are identical for all three models.  This is because each model is characterized by the same CFRg mass function, the same total mass in star-forming gas (thus the same CFRg mass function normalization) and the same final bound fraction $F_{bound}$ (since all simulations are performed with identical $SFE$, $\tau_{gexp}/\tau_{cross}$ and $r_h/r_t\lesssim 0.01$; see Eq.~\ref{eq:fb}).

\begin{figure}
\includegraphics[width=\linewidth]{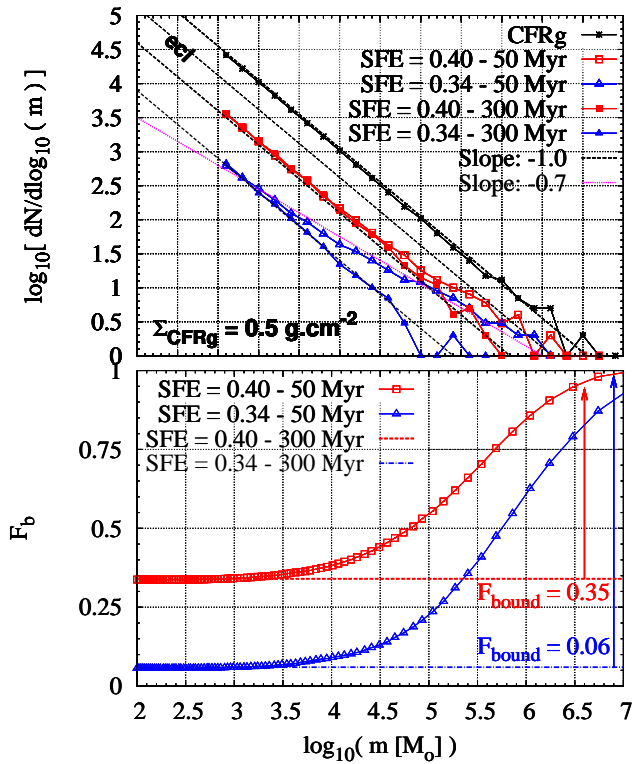}
\caption{{\it Top panel:} Cluster mass function in dependence of age and SFE (see key) for CFRgs with a constant mean surface density.  In addition to lessening the cluster survival rate, a lower SFE is conducive to a greater slope change during violent relaxation.  {\it Bottom panel:} Bound fraction $F_b$ in dependence of $m_{CFRg}$ for two ages and two SFEs.  At an age of 50\,Myr, a lower SFE stengthens the contrast between the bound fractions at low and high masses (compare the blue vertical arrow, $SFE=0.34$, to the red one,  $SFE=0.40$), hence the larger slope change at lower SFEs.       \label{fig:disc} }
\end{figure}

Figure \ref{fig:mfb} provides another perspective on the same processes.  It shows the instantaneous bound fraction $F_b$ as a function of $m_{CFRg}$ for the same ages as in Fig.~\ref{fig:MF}.  In other words, it represents on a linear scale the horizontal shift between each cluster mass function and the embedded-cluster mass function as a function of $m_{CFRg}$.  Panel sequence and colour/symbol codings are identical to Fig.~\ref{fig:MF}.  Figure \ref{fig:mfb} illustrates both the $F_b$ decrease with time (i.e. as infant-weight loss proceeds), and the mass-dependent behaviour of $F_b$ for the $\Sigma _{CFRg}$ and $r_{CFRg}$ models.  In each panel, the (black) horizontal dashed line marks the final bound fraction, $F_{bound}=0.11$ for our model parameters.  It is reached first by the high-mass clusters (mass $>10^5\,M_{\odot}$) in the $r_{CFRg}$ model, while the opposite is true for the $\Sigma _{CFRg}$ model.  The mass-dependent evolutionary rate is also highlighted at the bottom of each panel, where a time-span of 10\,Myr since gas expulsion is given in crossing-time units for CFRgs of masses $10^3\,M_{\odot}$, $10^5\,M_{\odot}$ and $10^7\,M_{\odot}$.  

The top panel of Fig.~\ref{fig:disc} illustrates the cluster mass function evolution for the $\Sigma_{CFRg}$ model with an $SFE$ lower ($SFE=0.34$) or higher ($SFE=0.40$) than in top panel of Fig.~\ref{fig:MF}.  At an age of $50$\,Myr, $SFE=0.34$ (blue line with open triangles) leads to a change of the cluster mass function slope greater than $SFE=0.40$ (red line with open squares).  The slope is $-0.7$ for $SFE=0.34$, while it stays close to $-1$ over most of the cluster mass range when $SFE=0.40$.  This behaviour is explained in the bottom panel which shows $F_b$ as a function of $m_{CFRg}$.  A lower $SFE$ leads to a smaller final bound fraction $F_{bound}$, thereby enhancing the contrast (see the vertical arrows) between high-mass clusters (mass $>10^6\,M_{\odot}$), which have just started to give off stars beyond their tidal radius) and low-mass ones (mass $<10^4\,M_{\odot}$, which have completed their violent relaxation by an age of 50\,Myr).  In turn, a lower $SFE$ strengthens the change of the cluster mass function slope. \\

\begin{figure}
\includegraphics[width=\linewidth]{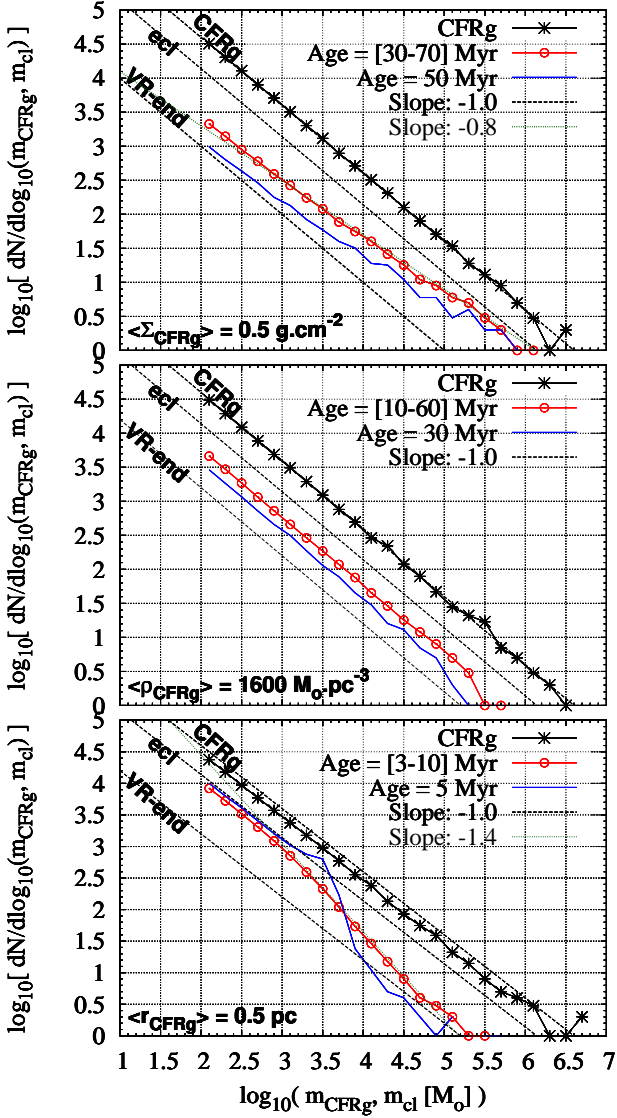}
\caption{Cluster mass functions (red line with open circles) integrated over a given age range (see key) typical of violent relaxation for three mean CFRg mass-radius relations: constant mean surface density ($<\Sigma_{CFRg}>$), constant mean volume density ($<\rho_{CFRg}>$) and constant radius ($<r_{CFRg}>$), from top to bottom, respectively.  Compared to Fig.~\ref{fig:MF}, the simulations now add a Gaussian noise to the mass-radius relations of Table~\ref{tbl:mrr} and include a Gaussian distribution of the SFE [see text for details].  For the sake of comparison with the first set of simulations presented in Fig.~\ref{fig:MF}, each panel shows the corresponding single-age mass function (solid blue lines) and the thick dashed black lines `CFRg', `ecl' and `VR-end'.  As previously, the mass of a cluster is defined as the stellar mass enclosed within its tidal radius at the age of relevance.  \label{fig:MFblur} }
\end{figure}

The models leading to the cluster mass functions presented in Fig.~\ref{fig:MF} build on a single variable, namely, the CFRg mass.  Other model parameters are either fixed (e.g. the cluster age, the CFRg star formation efficiency) or depend on the CFRg mass (e.g. the CFRg radius for the $\rho_{CFRg}$ and $\Sigma_{CFRg}$ models).  Yet, real star cluster systems probably form out of CFRgs with a range of SFEs and a mass-radius relation noisier than a pure power-law.  In addition, observed cluster mass functions are integrated over age ranges of several Myr to many tens of Myr, depending on the cluster mean age.  We now investigate how the cluster mass function patterns identified in Fig.~\ref{fig:MF} respond when the simulations include, simultaneously: {\it (i)} a cluster age range, {\it (ii)} a Gaussian distribution of the SFE and {\it (iii)} a power-law CFRg mass-radius relation superimposed with Gaussian noise.  We assume a Gaussian SFE distribution with a mean $\overline{SFE}=0.35$ and a standard deviation $\sigma_{SFE}=0.05$.  For each CFRg, a Gaussian noise is added to the logarithm of its radius predicted by the mass-radius relations used in Fig.~\ref{fig:MF}.  We adopt a standard deviation of $0.17$ in $log_{10}(r_{CFRg})$.  As a result, $>99$\% of the CFRgs have SFEs between 0.20 and 0.50 (i.e. the $6\sigma$-width of the Gaussian distribution), and span a  factor of 10 in radius around the mass-radius relations of Fig.~\ref{fig:mrr} and Table~\ref{tbl:mrr}.  

This second set of cluster mass functions is presented as the (red) lines with open circles in Fig.~\ref{fig:MFblur}.  The age ranges over which they are integrated are: 30-70\,Myr, 10-60\,Myr and 3-10\,Myr for the $\Sigma_{CFRg}$, $\rho_{CFRg}$ and $r_{CFRg}$ models, respectively.  For the sake of comparison with the first set of simulations, Fig.~\ref{fig:MFblur} also shows the corresponding single-age mass functions (thin solid blue lines) and the thick (black) dashed lines `CFRg', `ecl' and `VR-end' of Fig.~\ref{fig:MF}.  

Compared to the first set of simulations, the mass function features of the $r_{CFRg}$ model are weakened, with the three-segment mass function turned into a double-index power-law (compare the blue solid line and the red line with open circles in the bottom panel of Fig.~\ref{fig:MFblur}).  For masses lower than $10^3\,M_{\odot}$, the slope of the new cluster mass function mirrors that of the CFRgs while, at higher masses, it steepens up to $-1.4$.  Such a change should still be detectable in observed cluster data sets.  As for the $\Sigma_{CFRg}$ model, the cluster mass function slopes from the first and second simulation sets do not differ significantly (see the solid blue line and the red line with open circles in top panel of Fig.~\ref{fig:MFblur}).  Along the course of violent relaxation, the slope of the cluster mass function changes by $\simeq 0.2$ (from $-1$ to $-0.8$, then back to $-1$), which is probably just at the detectability limit.  

We stress again that none of our simulations include a significant external tidal field.  Yet, the tidal field impact may be important for the high-mass CFRgs of the $\Sigma_{CFRg}$ model.  We come back to this point later in this section.          

How much slope variation do observed cluster mass functions show?  The evolution with time of the cluster mass functions in the Solar Neighbourhood and in the Large Magellanic Cloud was investigated by \citet{pis08} and \citet{deg06}, respectively.  In the Large Magellanic Cloud, \citet{deg06} find that the cluster mass function steepens from $-0.80 \pm 0.10$ for clusters younger than 30\,Myr to $-0.98 \pm 0.08$ for clusters younger than 500\,Myr (their fig.~8 and table 3).  A similar trend -- albeit at younger ages -- was identified by \citet{pis08}.  They report a steepening of the Galactic open cluster mass function from $-0.66 \pm 0.14$ for ages younger than 8\,Myr to $-1.13 \pm 0.08$ for clusters younger than 300\,Myr (their fig.~6).  Although these two examples seem to point towards CFRgs of constant mean surface density, they both require strong cautionary notes.   
The cluster mass estimates inferred by \citet{pis08} are tidal masses, that is, they represent the total mass in stars enclosed within the cluster tidal radius, $r_t$.  As the cluster mass scales with $r_t^3$, any uncertainty on $r_t$ strongly affects the cluster mass.  This may explain why some of their cluster masses are higher ($3 \cdot 10^5M_{\odot}$) than the mass estimate of \citet{cla05} for Westerlund-1, the most massive Galactic cluster ($10^5M_{\odot}$).  
As for the Large Magellanic Cloud, at ages younger than 200\,Myr, \citet{bau1X} find a cluster mass function slope of $-1.32 \pm 0.11$, which is significantly steeper than that inferred by \citet{deg06} \citep[$-0.94 \pm 0.10$; see][for a full discussion]{bau1X}.  
Finally, we note that \citet{fal09} also find a slight evolution of the cluster mass function slope from $-1.14 \pm 0.03$ over the age range 1-10\,Myr, to $-1.03 \pm 0.07$ over the age range 10-100\,Myr.  This effect is hardly significant compared to the error bars, however.

In addition, we remind the reader that the results presented in Figs.~\ref{fig:MF}-\ref{fig:MFblur} assume a {\it weak tidal field impact} (i.e. $r_h/r_t \lesssim 0.01$).  The impact of a tidal field upon a population of CFRgs is tightly related to their mass-radius relation \citep[see][for a thorough discussion]{par11}.  Let us assume that all CFRgs are exposed to the same tidal field (e.g. they are all located in the same region of a given galaxy).  If the mean surface density of CFRgs is constant, more massive objects have a lower {\it volume} density and are therefore more vulnerable to tidally-induced mass-losses after gas expulsion.  If the tidal field is strong enough, 
a deficit in high-mass clusters then characterizes the end of violent relaxation (see bottom panel of fig.~3 in \citet{par11} and fig.~4 in \citet{par10}), and the cluster mass function becomes {\it steeper} than the embedded-cluster mass function.  In contrast, CFRgs with a given mean {\it volume} density all present the same sensitivity to the external tidal field,  thereby preventing a distorsion of the cluster mass function shape through mass-dependent tidally-induced mass-losses. 
  
The nature of the CFRg mass-radius relation remains heavily debated.  Based on an analysis of the binary population in young star clusters,  \citet{mar12} favour a weak mass-radius relation (i.e. close to a constant radius).  In contrast, \citet{fal10} infer a constant-surface-density relation based on the SFE -- hence the amount of stellar feedback -- required to cleanse a CFRg from its residual star-forming gas.  Note that their analysis does not explore the impact of a mass-dependent gas-expulsion time-scale or of a tidal field \citep[see section 4.1 in][for a detailed discussion of these points]{par11}.

We strongly encourage observers to report -- without any preconceived idea -- how the cluster mass function evolves with time, and to vary the size of the cluster age ranges over which cluster mass functions are integrated.  This also requests a careful assessment of the errors affecting the mass function slope and, therefore, a good control of the errors affecting individual cluster mass estimates.

\section{Returning to virial equilibrium: A radially dependent time-scale}
\label{sec:ap}
 
\subsection{From the cluster central regions to its outskirts: understanding the crucial effect of the aperture}
\label{ssec:ap1}

In the previous section, we estimated the time-scale over which infant weight-loss is completed in the case of a weak tidal field.  
That is, we assessed how long it takes for all due-to-be unbound stars to cross the cluster limiting tidal radius.  These time-scales are of order several tens of million years.  They    
depend on the CFRg crossing-time, and also on the tidal field strength: a stronger tidal field is conducive to a smaller tidal radius hence a shorter time-span for the unbound stars to cross it.  
The simulations of the previous section being performed under the assumption of a weak tidal field, the time-scales derived in Figs~\ref{fig:MF}-\ref{fig:MFblur} are upper limits.

The time-scale needed for the cluster {\it central} regions to return to virial equilibrium is even shorter.  This radial-dependence may contribute to explaining why several star clusters are reported to be in virial equilibrium despite being no older than a few Myr \citep[e.g. Westerlund-1, hereafter Wd~1; ][]{men09}.  Since star clusters are expected to be out of virial equilibrium after gas expulsion, it may be tempting to conclude from this type of observations that gas expulsion plays only a minor role in the early evolution of star clusters (through e.g. a high $SFE$, adiabatic gas expulsion ($\tau_{gexp} >> \tau_{cross}$), or a subvirial state: see \citet{goo09}).  In this section, we build on the observations of \citet{men07,men09} to demonstrate that this is not necessarily the case.  

\citet{men09} estimate the dynamical mass of Wd~1, $M_{dyn}$, based on the radial velocity dispersion of 10 massive stars.  They find a value comparable to the cluster luminous mass estimate, $M_{lum}$.  That is, Wd~1 is observed in virial equilibrium, a result which may seem unexpected for a 4-Myr old star cluster \citep{gen11} if it is born out of a gas expulsion process.  The discrepancy is lifted, however, when the high density of the Wd~1 CFRg -- hence its short crossing-time -- and the limited size of the observed field-of-view are accounted for.  In what follows, the observed half-light radius, $r_{hl} \simeq 0.86$\,pc, and luminous mass, $M_{lum} \simeq 10^5\,M_{\odot}$, of Wd~1 are taken from \citet{men09}.  As for the age, we use the 4-Myr estimate of \citet{gen11} rather than the 6-Myr estimate of \citet{men09}.  This equates with a time-span of 3\,Myr after gas expulsion if gas dispersal took place when the cluster was $\simeq 1$\,Myr old.  Here, it is crucial to keep in mind that the return to virial equilibrium of the cluster does {\it not} depend on the physical time-span since gas expulsion (i.e. the time expressed in units of Myr).  Rather, it depends on the time-span expressed in units of the {\it initial} crossing-time, namely, the CFRg crossing-time.  A cluster can therefore be dynamically old despite a young stellar age.

The 10 stars used to derive the radial velocity dispersion are all observed at projected distances less than 2.5\,pc from the cluster centre \citep[see both figs.~1 in][]{men07,men09}.  Assuming a Plummer profile, the cluster virial radius, $r_v$, follows from the projected half-light radius $r_{hl}$: $r_v = 1.3r_{hm} \simeq 1.5$\,pc, with $r_{hm}$ the 3-dimensional half-mass radius, and $r_{hm}=1.3r_{hl}$, where the factor 1.3 accounts for the projection onto the sky.  With $M_{lum} \simeq 10^5\,M_{\odot}$, the {\it present-day} crossing-time at the virial radius \citep[eq.~6 in][]{bau07,heg03} of Wd~1 is $\tau_{cross} \simeq 0.25$\,Myr.
\footnote{A similar value characterizes NGC~3603 with $m_{cl} \simeq 10^4\,M_{\odot}$ and $r_{hm} \simeq 0.5$\,pc \citep{roc10}}  
As we shall see in Fig.~\ref{fig:rhl}, the observed half-light radius is aperture-dependent, an effect which complicates further the estimate of the initial crossing-time.

\begin{figure}
\includegraphics[width=\linewidth]{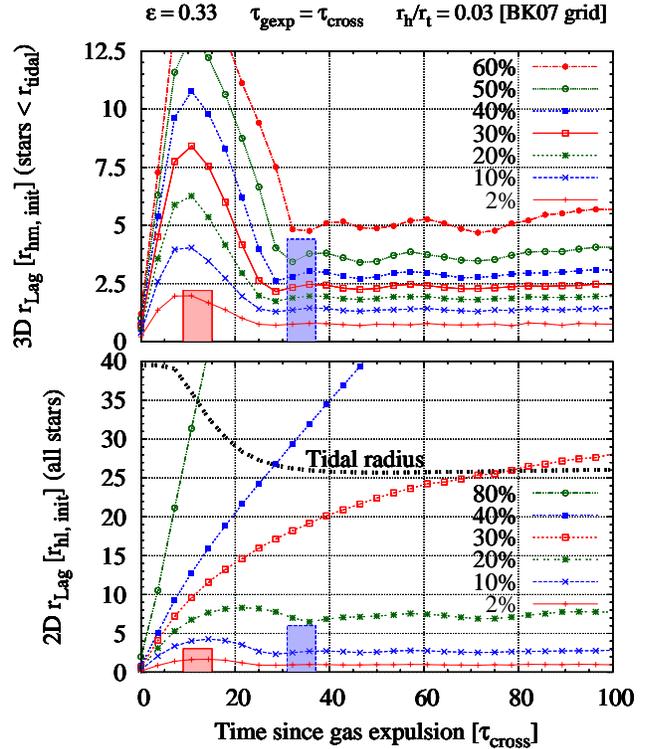}
\caption{{\it Top panel:} Evolution of the 3-dimensional Lagrangian radii of stars within the instantaneous tidal radius (depicted as the symbol-free dotted black line in the bottom panel) for one of the model clusters of \citet{bau07}.  SFE, gas expulsion time-scale and tidal field impact are quoted on top of the panel.  {\it Bottom panel:} Evolution of the projected (2-dimensional) Lagrangian radii of all stars (bound and unbound).  In both panels, mass fractions are quoted in the key.  Filled rectangles represent two evolutionary stages of Westerlund-1 since gas expulsion ($x$-axis), and the radius of the field-of-view encompassing the stars observed by \citet{men09} ($y$-axis) (see text for details).           
\label{fig:ap} }
\end{figure}

Figure \ref{fig:ap} depicts the time-evolution of the Lagrangian radii of a model star cluster with initial parameters: SFE=0.33, $r_h/r_t=0.03$ and $\tau_{M}=0.33\tau_{cross}$, where $\tau_M$ is the gas expulsion time-scale as defined by eq.~1 of \citet{bau07}.  
Note that \citet{bau07} model gas expulsion as an exponential decrease with time of the cluster gas content.  Their gas expulsion time-scale $\tau _{M}$ corresponds to the time when a fraction $e^{-1}=0.37$ of the initial gas mass is left.  We define $\tau_{gexp}=3\tau_{M}$ so that $\tau_{gexp}$ corresponds to a residual gas mass fraction of $e^{-3}=0.05$, i.e., the cluster is practically devoid of gas.  In Fig.~\ref{fig:ap}, we thus have $\tau_{gexp} =\,\tau_{cross}$.
The top panel shows the 3-dimensional Lagrangian radii of stars within the instantaneous tidal radius.  The bottom panel represents the projected Lagrangian radii of {\it all} stars, bound and unbound (i.e. as would be seen by an observer).  Units of the $x$-axis, top and bottom $y$-axes are the initial crossing-time, initial half-mass radius and initial half-light radius, respectively.  In a first step, let us approximate the initial crossing-time and initial half-light radius with their observed estimates.  The cluster evolutionary stage after gas expulsion ($3\,Myr \equiv 12\,\tau_{cross}$ if $\tau_{cross}=0.25$\,Myr), and the radius of the field-of-view containing the stars used by \citet{men09} to derive the radial velocity dispersion (2.5\,pc/0.86\,pc=2.9$r_{hl}$, or $2.3\,r_{hm}$), are highlighted as the left (red) filled rectangles in both panels of Fig.~\ref{fig:ap}.  The targeted region of the cluster is still out of equilibrium
\footnote{Note that the maximum reached by the 10\,\% and 20\,\% projected Lagrangian radii at times of 15-20\,$\tau_{cross}$ (bottom panel of Fig.~\ref{fig:ap}) does not imply the absence of radial motions, but the simultaneous contraction of the inner regions and expansion of the outer regions}.  
However, a cluster expands and loses mass following gas expulsion.  Its {\it initial} crossing-time is thus shorter than presently observed and the cluster is now at an accordingly older dynamical stage.  Let us make the conservative assumption that the cluster has expanded by a factor 2 only, and let us neglect mass-loss.  This implies $y$- and $x$-axis units smaller by factors 2 and $2\sqrt{2}$, respectively.  The rectangle depicting the parameter space covered by \citet{men09} is thus shifted towards older evolutionary stages by a factor 2.8, while its height is increased by a factor 2 (right blue rectangles).  It now appears that for the inner regions of Wd~1 covered by the field-of-view of \citet{men09}, the response to gas expulsion may just be over.  We analyse in greater detail the case of Wd~1 in Sects \ref{ssec:ap2} and \ref{ssec:ap3}, where we demonstrate that the hypothesis of virial equilibrium despite a significant gas expulsion episode some 3\,Myr ago is indeed viable.

Coeval with the central regions which have returned to equilibrium, outer layers keep expanding, thereby bringing more stars accross the tidal radius (thick dotted line in bottom panel).  
The central regions of a cluster may thus be observed in virial equilibrium while, on a larger spatial scale, its external layers keep expanding.  This explains straightforwardly why some Galactic massive star clusters as young as a few Myr are observed in virial equilibrium (e.g. Wd~1), while clusters with ages 10-60\,Myr still show a perturbed surface brightness profile at radii $>10$\,pc \citep[e.g M82-F, NGC1569-A and NGC1705-1; see fig.~1 in][]{bas06}.  Owing to their distance, clusters in starburst galaxies are often observed with large linear apertures, thereby highlighting the shells of unbound stars expanding on spatial scales $>10$\,pc.  In contrast, observations of star clusters in the Galactic disc are often limited to their few central parsecs (see Sects~\ref{ssec:ap2} and \ref{ssec:arch}).  We now quantify how the size of the aperture affects the observed ratio of the dynamical and luminous masses of a cluster, $M_{dyn}/M_{lum}$.  In other words, we derive $M_{dyn}/M_{lum}$ as would be done by an observer.   

\subsection{Quantifying the impact of the aperture}
\label{ssec:ap2}

When the aperture is large, the cylinder it defines (i.e. the cylinder with the line-of-sight as its main axis) intercepts a greater fraction of the outer shells containing the stars expanding in the vicinity of the tidal radius and beyond.  As a result, the measured radial velocity dispersion rises. 
This simple exercise demonstrates the importance of the size of the field-of-view: the smaller the aperture, the smaller the observed departure from virial equilibrium, and the shorter the time-span required for the cluster targeted region to return to equilibrium.       

\begin{figure}
\includegraphics[width=\linewidth]{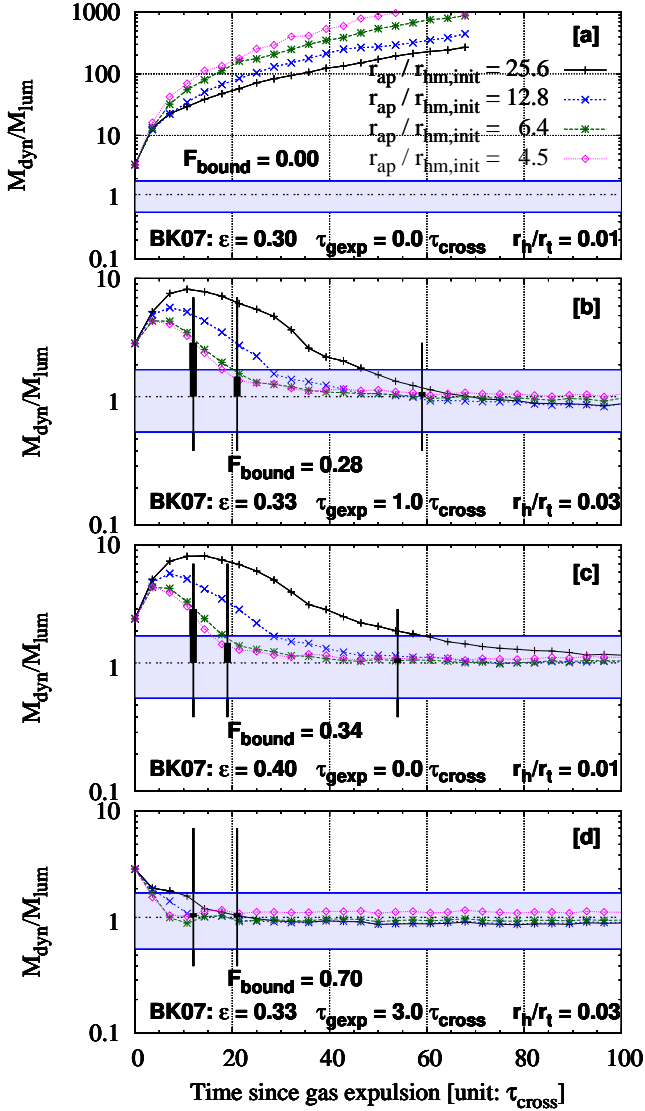}
\caption{
Evolution with time of the cluster dynamical-to-luminous mass ratio, $M_{dyn}/M_{lum}$, for the stars contained in an aperture of radius $r_{ap}$ (based on the $N$-body simulations of \citet{bau07}).  The dynamical mass, $M_{dyn}$, stems from the radial velocity dispersion (Eq.~\ref{eq:Mdyn_rap}), not from star proper motion measurements.  The size of the aperture is expressed in units of the cluster initial half-mass radius, $r_{hm, init}$ (see key in top panel).  Gas-expulsion model parameters are quoted at the bottom of each panel, along with their final bound fraction $F_{bound}$.  Note that the gas-expulsion time-scale $\tau_{gexp}$ is defined as $\tau_{gexp} = 3\tau_M$, where $\tau_M$ is the e-folding time of gas expulsion of \citet[][their eq.~1]{bau07}.  The vertical line segments correspond to evolutionary stages of Wd~1 based on various initial crossing-time estimates and a time-span of 3\,Myr since gas expulsion (see text for details).  The shaded area illustrates the range of uncertainties affecting $M_{dyn}$ in \citet{men07, men09} 's studies.  Note that the $y$-range in panel [a] is much larger than in panels [b-d]. 
\label{fig:mdml} }
\end{figure}

Figure \ref{fig:mdml} shows the post-gas-expulsion evolution of $M_{dyn}/M_{lum}$.
Each panel corresponds to a different set of model parameters, i.e. star formation efficiency ($\epsilon$), gas expulsion time-scale ($\tau_{gexp}/\tau_{cross}$), tidal field impact ($r_h/r_t$) and their corresponding final bound fraction ($F_{bound}$).  The cases covered by Fig.~\ref{fig:mdml} range from the complete disruption of the cluster ($F_{bound}=0$ in top panel [a]) to a cluster barely affected by gas expulsion ($F_{bound} \lesssim 1$ in bottom panel [d]).
Note the degeneracy between panels [b] and [c]: the lower SFE of panel [b] (SFE=0.33 compared to 0.40 in panel [c]) is compensated by its longer gas-expulsion time-scale ($\tau_{gexp}=\tau_{cross}$ compared to $\tau_{gexp}=0$ in panel [c]).  Each panel considers in turn four aperture radii, $r_{ap}=$ 3.5, 5, 10 and 20\,pc, which correspond to 4.5, 6.4, 12.8 and 25.6 cluster initial half-mass radii ($r_{hm, init}=0.8$\,pc; see key in top panel).  Measuring in our simulations the observed half-light radius for apertures smaller than 3.5\,pc proved inaccurate and results for smaller apertures are therefore not presented here.  With the aperture and time expressed in units of $r_{hm,init}$ and $\tau_{cross}$, respectively, the results of Fig.~\ref{fig:mdml} can be rescaled to clusters with different initial densities, sizes, and apertures.     

The luminous mass $M_{lum}$ is the total (true) mass contained within the aperture.  The cluster dynamical mass $M_{dyn}$ is derived from the radial velocity dispersion, $\sigma_{los}^2$, measured for all the stars contained within the aperture.  
Assuming a Plummer density profile, $M_{dyn}$, $\sigma_{los}^2$ and $r_{ap}$ are related through: 

\begin{equation}
M_{dyn}=\frac{32}{\pi} \cdot \frac{ \sigma_{los}^2 \cdot r_{hp} }{G}\cdot f(t) \simeq 10 \cdot \frac{\sigma_{los}^2 \cdot r_{hp}}{G} \cdot \frac{t^2 \cdot \sqrt{1+ t^2} }{(1+t^2)^{1.5}-1}\,,
\label{eq:Mdyn_rap}
\end{equation}

where $r_{hp}$ is the projected half-mass radius observed within the aperture (equivalent to the half-light radius if light follows mass, and to the Plummer radius for a Plummer model), and $t=r_{ap}/r_{hp}$.  For large apertures, i.e. $r_{ap} >> r_{hp}$, Eq.~\ref{eq:Mdyn_rap} is reduced to the `canonical' form \citep[see eq.~3 in][]{fle06}:

\begin{equation}
M_{dyn} \simeq 10 \frac{\sigma_{los}^2 \cdot r_{hp}}{G}\;.
\end{equation}    

As anticipated above, when the aperture is larger, $M_{dyn}/M_{lum}$ is higher and it takes longer for the observed cluster region to recover $M_{dyn}/M_{lum} \simeq 1$  (compare the black curves with plus-signs to e.g. the pink curves with open diamonds in panels [b-d] of Fig.~\ref{fig:mdml}).  

\begin{figure}
\includegraphics[width=\linewidth]{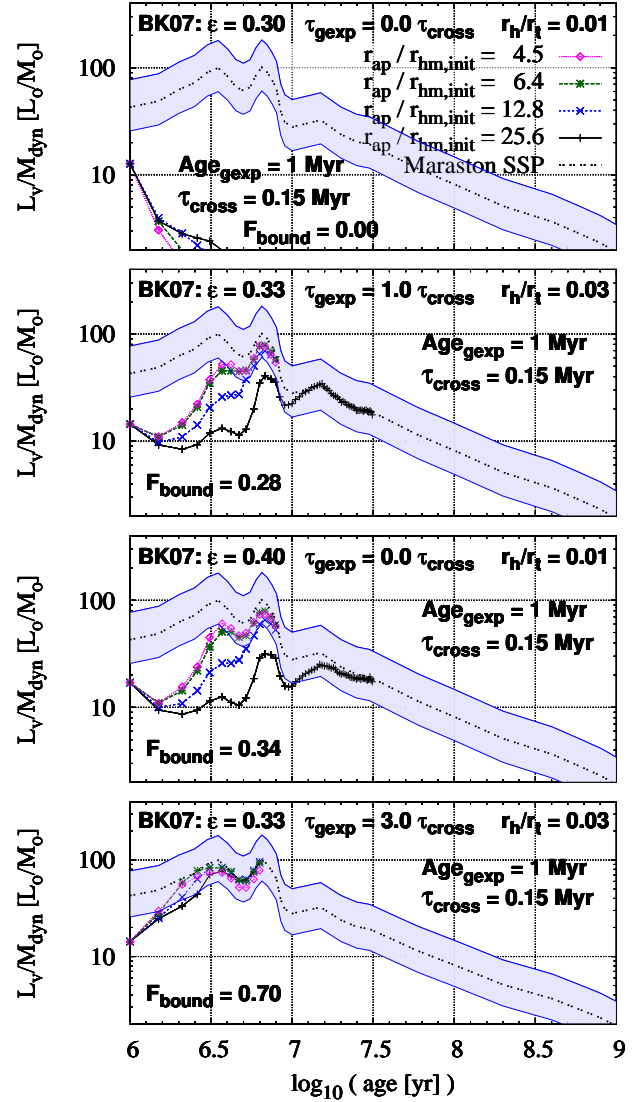}
\caption{
Evolution of the ratio between the visual luminosity, $L_v$, of a star cluster and its dynamical mass, $M_{dyn}$, with both quantities measured over a cylindrical aperture of radius $r_{ap}$.  These diagrams are the counterparts of those in Fig.~\ref{fig:mdml} (identical colour/symbol codings).  That is, the time evolution of $M_{dyn}/M_{lum}$ has been combined with the age-dependent mass-to-light ratio of a single age stellar population according to \citet{mar05}.  We assume that gas is expelled at an age of 1\,Myr and that the initial crossing-time is 0.15\,Myr.  Note that time in this figure is measured in years, while the time-unit in Fig.~\ref{fig:mdml} is the cluster initial crossing-time.       
\label{fig:mdlv} }
\end{figure}
    
The vertical extent of the (blue) shaded area in Fig.~\ref{fig:mdml} illustrates the uncertainty range affecting the Wd~1 dynamical mass estimate, hence the ratio $M_{dyn}/M_{lum}$.     
\citet{men07} and \citet{men09} estimate $M_{dyn}$ to be $0.63_{-0.37}^{+0.53} \times 10^5\,M_{\odot}$ and $1.5_{-0.7}^{+0.9} \times 10^5\,M_{\odot}$, respectively.  Because the accuracy of the first estimate is hindered by the small number (4) of stars studied, \citet{men09} increased their sample up to 10 stars.  They stress, however, that this larger sample includes 5 (presumed) yellow hypergiants whose radial pulsations may contribute an increase of the measured velocity dispersion.  Either estimate thus suffers from its own drawback.  In what follows, we adopt the average of both values, namely, $1.1_{-0.5}^{+0.7} \times 10^5\,M_{\odot}$.  As for the cluster luminous mass, $M_{lum} \simeq 10^5\,M_{\odot}$ \citep{cla05, men07}, we ignore the uncertainties affecting it.   
The  $M_{dyn}/M_{lum}$ ratio is thus $1.1_{-0.5}^{+0.7}$ and is depicted as the (blue) shaded area in Fig.~\ref{fig:mdml}.  
The vertical (black) line segments to the left of panels [b-d] indicate the evolutionary stage of Wd~1 after gas expulsion if the initial and observed crossing-times are equal, i.e. as for the red (leftward) rectangles in Fig.~\ref{fig:ap}: $t=3\,Myr \equiv 12\tau_{cross}$.  The line segment thick part indicates the difference between the observations (i.e. the horizontal dotted line at $M_{dyn}/M_{lum}=1.1$) and the models with $r_{ap}/r_{hm,init}$=4.5 and 6.4 (see Sect.~\ref{ssec:ap3}).

\begin{figure}
\includegraphics[width=\linewidth]{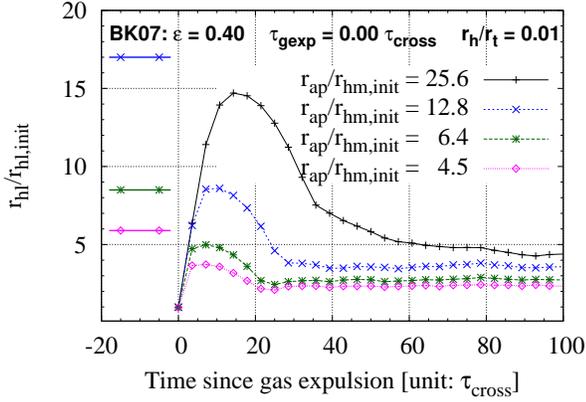}
\caption{
Evolution of the (projected) half-light radius observed within different apertures for the quoted SFE $\epsilon$, gas-expulsion time-scale $\tau_{gexp}$ and tidal field impact $r_h/r_t$.  The aperture sizes in units of the initial half-mass radius, $r_{ap}/r_{hm, init}$, are given in the key.  In the left part of the panel, the horizontal line segments indicate the three smallest apertures in units of the initial half-light radius, $r_{ap}/r_{hl, init}$.  For the $N$-body model of \citet{bau07} under scrutiny here, $r_{hl, init} \simeq 0.6$\,pc and $r_{ap}=$3.5, 5, 10 and 20\,pc which gives  $r_{ap}/r_{hl, init}=$5.9, 8.5, 17 and 34.  The last and largest aperture is beyond the plot top border.      
\label{fig:rhl} }
\end{figure}

The amount of dynamical evolution experienced by a cluster after gas expulsion depends on its {\it initial} crossing-time, namely, the crossing-time of the CFRg out of which it formed and defined by Eq.~\ref{eq:tcr}.  The observed crossing-time of a gas-free cluster is longer than that of its CFRg because of {\it (i)} the gas loss, {\it (ii)} the infant weight-loss, and {\it (iii)} the gas-expulsion-driven cluster expansion.  In other words, using the observed crossing-time leads us to underestimate the cluster dynamical age, an effect introduced in Fig.~\ref{fig:ap}.  
In addition, a consistent comparison between the models and the observations requires the size of the aperture {\it in units of the initial half-mass radius}, $r_{ap}/r_{hm,init}$, which is not accessible to observations.  We now derive estimates of the initial crossing-time and initial half-mass radius.

\subsection{An attempt to recover the initial crossing-time}
\label{ssec:ap3}

The radius of the field of view of Wd~1 in \citet{men07} and \citet{men09} is $\simeq 2.5$\,pc.  
Since the initial half-mass radius is unknown, we resort to rough estimates of the ratio $r_{ap}/r_{hm,init}$ in the first place.  We assume $0.3\,pc \lesssim r_{hm,init} \lesssim 1$\,pc which leads to $8 \gtrsim r_{ap}/r_{hm,init} \gtrsim 2.5$.  In Fig.~\ref{fig:mdml}, we therefore consider models with $r_{ap}/r_{hm,init}$=4.5 and 6.4.   

To answer the question as to whether Wd~1 can be observed in virial equilibrium despite a major gas-expulsion phase some 3\,Myr ago, we focus on panels [b] and [c] of Fig.~\ref{fig:mdml} where the final infant weight-loss is significant, i.e. two-thirds of the cluster initial stellar mass are lost.  (Panel [a] presents the case of a fully disrupted cluster, irrelevant here, and we come back to panel [d] later in the discussion).  
Building on the {\it observed} crossing-time, we find a discrepancy between the models (green curve with asterisks and pink curve with diamonds) and the observations (horizontal black dotted line) larger than 3$\sigma$, as quantified by the thick part of the most leftward vertical segment.  As we saw in Fig.~\ref{fig:ap}, however, using the observed crossing-time underestimates the cluster evolutionary stage.  
So let us now correct the observed crossing-time for the gas-loss (by multiplying it by $SFE^{1/2}$; see Eq.~\ref{eq:tcr}).  This leads to the cluster evolutionary stages indicated by the second vertical segment from left in panels [b-c] at $t \simeq 20\tau_{cross}$.  This sole correction brings the gas-expulsion models within 1$\sigma$ of the observations when $r_{ap} < 7 r_{hm, init}$.  
In addition to gas loss, the cluster is more compact prior to gas expulsion.      
Figure \ref{fig:rhl} illustrates the evolution of the observed projected half-light radius for the quoted SFE, gas-expulsion time-scale and tidal field impact (same parameters as in panel [c] of Fig.~\ref{fig:mdml} and model outputs similar to panel [b]).  
The {\it observed} half-light radius is -- in essence -- aperture-limited.
The apertures used to observe the model cluster are shown as horizontal lines in the left part of the plot with identical colour/symbol codings.  Note that they have been rescaled in units of the initial half-light radius and that the largest aperture is out of range.  For the apertures and dynamical evolutionary stages of relevance here ($8 \gtrsim r_{ap}/r_{hm,init} \gtrsim 2.5$, $t > 20\tau_{cross}$ after correcting for the gas loss), the cluster has expanded by a factor of $\simeq 2$.  The initial half-light radius may thus have been $r_{hl,init} \simeq 0.43$\,pc \citep[half the value measured by ][]{men07} which, for a Plummer profile, corresponds to an initial half-mass radius $r_{hm,init} \simeq 0.56$\,pc (since $r_{hm,init} =1.3 r_{hl,init}$).
The factor of two in expansion suggests that the initial crossing-time may have been as short as $\tau_{cross} \simeq 0.25\,Myr \times SFE^{1/2} \times (1/2)^{3/2} \simeq 0.06$\,Myr.  The corresponding dynamical ages after gas expulsion are shown as the vertical line segments at $t/\tau_{cross} \simeq 55-60$ in panels b-c of Fig.~\ref{fig:mdml}.  The observed $M_{dyn}/M_{lum}$ ratio and inferred evolutionary stages $t/\tau_{cross}$ of Wd~1 agree neatly with a gas expulsion scenario in which the stellar mass of the cluster after revirialization amounts to one-tenth only of the CFRg mass (since $\epsilon \simeq 0.33$ and $F_{bound} \simeq 0.33$).  The $\simeq 70$\%- infant weight-loss is visible as the decrease with time of the cluster tidal radius in the bottom panel of Fig.~\ref{fig:ap}: $r_t \propto (m_{cl})^{1/3} \propto (F_b \times m_{ecl})^{1/3}$, or an eventual decrease of $F_{bound}^{1/3} = 0.30^{1/3}$.
It is therefore worth emphasizing that star clusters with {\it stellar} ages of several Myr, which formed out of dense molecular clumps with short crossing-times, are already dynamically evolved, that is, the time-span since gas expulsion can be many initial crossing-times.     

The value inferred for the initial crossing-time at the virial radius of Wd~1, $\tau_{cross} \simeq 0.06$\,Myr, may seem short.  Still, it is not too dissimilar from the crossing-time of the molecular clump G0.253+0.016 for which \citet{lon12} find $\tau_{cross} \simeq 0.17$\,Myr (their table 2).  We note that the mass of G0.253+0.016 and Wd~1 are similar (i.e. $\simeq 10^5\,M_{\odot}$).  If gas-expulsion played a major role in the early evolution of Wd~1, then Wd~1 formed out of a CFRg even more massive than G0.253+0.016.   
With $\epsilon \simeq 0.33$ and $F_{bound} \simeq 0.33$, the parent CFRg of Wd~1 may have been as massive as $10^6\,M_{\odot}$.  A crossing-time of $\tau_{cross} \simeq 0.06$\,Myr is then achieved if $r_{CFRg} \simeq 1.5$\,pc (see Eq.~\ref{eq:tcr} and Fig.~\ref{fig:mrr}).  This is reminiscent of the radius put forward by \citet{kro02} and \citet{kro05}, who assume that the CFRg radius is about 1\,pc regardless of the CFRg mass.  We now see a connection between the time-evolution of the cluster mass function shape (Sect.~\ref{sec:mod}) and how quickly star clusters return to virial equilibrium, the link between both aspects being the CFRg mass-radius relation.

What if the mass-radius relation of CFRgs is not one of constant radius?
If CFRgs have a constant mean volume density \citep{wu05, par11c}, then, at high mass, the field-of-view may not be large enough to cover the CFRg spatial extent in full.  For instance, if the mean volume density is that adopted in Fig.~\ref{fig:mrr} (i.e. $n_{H2} \simeq 2.3 \cdot 10^4\,cm^{-3}$), a CFRg $10^6\,M_{\odot}$ in mass now has a radius $r_{CFRg} \simeq 5$\,pc and a crossing-time $\tau_{cross} \simeq 0.4$\,Myr.  The dynamical evolution is thus markedly slower than the case detailed above for $r_{CFRg} \simeq 1.5$\,pc as a time-span of 3\,Myr after gas expulsion now corresponds to $\simeq 8\,\tau_{cross}$ only.  Yet, that does not necessarily imply that the cluster cannot be observed as having returned to equilibrium.   
The aperture used to observe Wd~1 ($\simeq 2.5$\,pc) is now twice as small as the size of our putative CFRg.  This limits the observations to the inner half of the parent CFRg, with two immediate  consequences.  Firstly, since molecular clumps have density gradients, the initial crossing-time of the observed field-of-view is shorter than the `global' value ($\tau_{cross} \simeq 0.4$\,Myr) quoted above.  Secondly, assuming a slope of $-1.7$ to $-2$ for the power-law density gradients of molecular clumps \citep{mue02}, we get $r_{ap} \simeq r_{hm,init}$ (that is, the aperture corresponds to about half of the three-dimensional CFRg mass), a case not investigated in Fig.~\ref{fig:mdml}.  
New models where the aperture covers a fraction only of the CFRg are therefore needed, a study which we defer to a future paper.

Another possibility is of course that gas expulsion is a minor perturbation to the cluster and the final bound fraction $F_{bound}$ is close to unity.  This is the case for instance because gas was expelled on an adiabatic time-scale as in panel [d] of Fig.~\ref{fig:mdml}.  The two vertical line segments depict the evolutionary stages of Wd~1 based on the observed and gas-loss-corrected crossing-times.  In case of adiabatic gas expulsion, the embedded cluster experience neither significant infant weight-loss nor significant spatial expansion and further corrections of the observed crossing-time are therefore superfluous.  
The limited cluster expansion allows the cluster to return to equilibrium on a time-scale as short as $\simeq 8\tau_{cross}$ for apertures smaller than $10r_{hm,init}$ (and about $3\tau_{cross}$ at the 1$\sigma$ level).    

To estimate how quickly the intra-cluster gas is removed from an embedded cluster is a complex problem.  If the ionized material simply expands at the sound speed, $vs \simeq 10\,km\cdot s^{-1}$, then a CFRg of mass $m_{CFRg} \simeq 10^6\,M_{\odot}$ and radius $r_{CFRg} \simeq 1.5$\,pc loses its gas on a time-scale of 0.15\,Myr, equivalent to a few crossing-times (since $\tau_{cross} \simeq 0.06$\,Myr; see above).  Gas expulsion is thus adiabatic \citep{kro02} and the return to equilibrium is best-described by panel [d].  The gas-expulsion time-scale can be shorter ($\tau_{gexp} \lesssim \tau_{cross}$), however, if radiation pressure dominates the gas-expulsion process \citep{kru09} and the young cluster dynamical evolution then follows from panels [b-c].

In Fig.~\ref{fig:mdml}, the vertical extent of the shaded (blue) area ignores the uncertainties affecting $M_{lum}$.  Were they included, the error on $M_{dyn}/M_{lum}$ would be accordingly larger.  In that respect, \citet{gen11} derive $M_{lum} \simeq 0.49_{-0.05}^{+0.18} \times 10^5\,M_{\odot}$ for Wd~1 \citep[a similar value was found by ][]{bra08}.  Combined to the $M_{dyn}$ estimate of \citet{men09}, this is suggestive of a higher ratio $M_{dyn}/M_{lum} \simeq 2$ (i.e. in Fig.~\ref{fig:mdml}, the shaded areas rise by a factor of $\simeq 2$), even more compatible with explosive gas expulsion conditions.  
Here lies one more source of complication, however.  In models, the stars used to derive the radial velocity dispersion $\sigma_{los}^2$ and the luminous mass estimate $M_{lum}$ are identical.  In real observations, that may not be the case.  To infer $M_{lum}$ and $M_{dyn}$ based on different observation sets, hence different fields-of-view, can only increase the error in $M_{dyn}/M_{lum}$.       

Figure \ref{fig:mdlv} combines the time-evolution of the $M_{dyn}/M_{lum}$ ratio of Fig.~\ref{fig:mdml} with the age-dependent mass-to-light ratio of a Simple Stellar Population as predicted by \citet{mar05} for a solar metallicity.  We assume that gas expulsion occurs when the cluster is 1\,Myr old.  Because the dynamical and photometric evolutions of star clusters have different time-units (initial crossing-time and Myr, respectively), Fig.~\ref{fig:mdlv} {\it demands} a reliable estimate of the initial crossing-time.  We take $\tau_{cross} = 0.15$\,Myr, a value similar to that inferred by \citet{lon12} for the molecular clump G0.253+0.016.  $\tau_{cross} = 0.15$\,Myr is also intermediate between our estimate of the initial crossing-time of Wd~1 and its observed value ($0.06$\,Myr and $0.25$\,Myr, respectively).  Panels, shaded areas and evolutionary tracks in Fig.~\ref{fig:mdlv} are the counterparts of those in Fig.~\ref{fig:mdml} (identical colour/symbol codings).

Some models of the time-evolution of $M_{dyn}/M_{lum}$ and $L_v/M_{dyn}$ have been presented in \citet{goo06}.  
Their simulations build on a Plummer model with a Plummer radius $r_P=3.5$\,pc and a star+gas mass $m_{CFRg} \simeq 5 \times 10^4\,M_{\odot}$.  This yields a virial radius of $r_v \simeq 1.7 \cdot r_P \simeq 5.8$\,pc and a crossing-time $\tau_{cross}=2.6$\,Myr \citep[using eq.~6 in][]{bau07}.  Compared to the initial crossing-time of a starburst cluster, this is about an order of magnitude too long.  We also note that their dynamical mass is measured from the velocity dispersion of stars located within 20\,pc of the cluster centre (equivalent to $4.5r_{hm,init}$ in their model).  In contrast, our velocity dispersion builds on the  stars contained within a cylindrical aperture.  Therefore, given that {\it (i)} it is unclear whether \citet{goo06} rescaled their model to a crossing-time appropriate for starburst clusters, and {\it (ii)} we define the $M_{dyn}/M_{lum}$ ratio based on different ensembles of stars, our results cannot be consistently compared to theirs.  Additionally, the tidal field strength is set to zero in their simulations, while Figs~\ref{fig:mdml}-\ref{fig:mdlv} builds on a weak tidal field.

\subsection{The Arches cluster: proper motions and extreme aperture}
\label{ssec:arch}

\begin{figure}
\includegraphics[width=\linewidth]{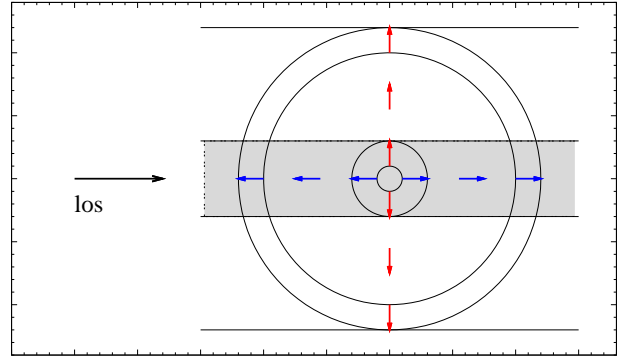}
\caption{Mapping the gas-expulsion driven expansion of a cluster.  The line-of-sight is indicated by the (black) arrow on the left.   Radial and proper motions are depicted by the horizontal (blue) and vertical (red) arrows.  The aperture is shown as the shaded area.  Because of the narrow centrally-located aperture, a proper motion mapping within the aperture misses the expansion of the cluster outer layers.  In contrast, radial velocities trace the whole radial extent of the cluster, from the central regions to the outskirts.  As a result, the time-evolutions of the dynamical masses inside the aperture inferred by proper motions on the one hand and radial velocities on the other hand are expected to differ.   
\label{fig:pm} }
\end{figure}

A recent study of the Arches cluster by \citet{cla12} infers estimates for its dynamical and luminous masses which are similar.  Because the Arches cluster is younger than Wd~1 \citep[age: 2-2.5\,Myr, ][]{naj04}, the time-span since gas expulsion may be as short as 10 initial crossing-times (assuming an age of 1\,Myr at gas expulsion and $\tau_{cross} \simeq 0.1$\,Myr).  At first glance, this seems to disprove a gas-expulsion scenario like that shown in panels b-c of Fig.~\ref{fig:mdml}.  However, it proves crucial to consider how the dynamical mass is derived and, again, how small the aperture is.

\citet{cla12} map proper motions of stars within the central 10\arcsec $\times$ 10\arcsec of the cluster (or 0.4\,pc $\times$ 0.4\,pc, equivalent to a projected radius of 0.2\,pc).  Combining their proper motion measurements with cluster modelling, they estimate  the dynamical mass contained within a cylinder of radius 0.4\,pc to be about $0.9 \cdot 10^4\,M_{\odot}$.  This is comparable to their luminous mass estimate\footnote
{Their photometric mass estimate is derived under the assumption of a top-heavy initial stellar mass function, an issue which remains disputed for the Arches cluster \citep[see e.g.][]{esp09}, and which is beyond the scope of the present discussion.}
within the same aperture, namely, $M_{lum} = $1-2$\cdot 10^4\,M_{\odot}$.  The small radius of 0.4\,pc is  comparable to, or smaller than, the initial half-mass radius of the cluster (see our estimate for Wd~1 above).  One thus needs a model with $r_{ap}/r_{hm,init} \lesssim 1$, a limit which our present $N$-body simulations do not achieve.  All we can anticipate is that, within such a small aperture, the cluster is observed to return to virial equilibrium even faster than in any of our simulations.  There is a strong caveat, however.  In the models presented in Fig.~\ref{fig:mdml}, the dynamical mass is estimated based on {\it radial velocity measurements} (see Eq.~\ref{eq:Mdyn_rap}).  The Arches cluster study of \citet{cla12} builds on {\it proper motion measurements}.  Although the development of a model akin to Eq.~\ref{eq:Mdyn_rap} for proper-motion-based dynamical masses is beyond the scope of the present paper, we point out that the measurements of proper motions in a small aperture  {\it necessarily} miss the expanding external layers which characterize a star cluster after gas expulsion.  This is illustrated in Fig.~\ref{fig:pm} where the horizontal (blue) arrows depict the radial motions and the vertical (red) ones depict the proper motions.  The aperture is shown as the shaded area.  While radial velocity measurements probe the full radial extent of the expanding cluster, proper motion measurements are restricted to the 3-dimensional central regions.  One may thus expect a lower measured velocity dispersion when based on proper motions than when based on radial velocities, hence $M_{dyn}^{pm} < M_{dyn}^{rv}$ where $M_{dyn}^{pm}$ and $M_{dyn}^{rv}$ are the dynamical mass estimates obtained from the proper motions and radial velocities, respectively.  To sum up, because of an aperture comparable in size to the initial half-mass radius of the cluster, and because of a cluster dynamical mass estimated from the proper motions of the cluster central regions only, it cannot be excluded that $M_{dyn}^{pm} \simeq M_{lum}$ is observed at the present Arches-cluster age even though the cluster experienced a major gas expulsion episode in the recent past.
   
\subsection{What must be kept on top of mind}
\label{ssec:sum}

The above discussions highlight the complexity of inferring whether the observed virial equilibrium of young gas-free star clusters is compatible with a gas-expulsion scenario.  A key-issue is that pivotal model parameters such as
{\it (1)} the initial crossing-time, hence the CFRg mean volume density, and 
{\it (2)} the size of the field-of-view with respect to the cluster initial half-mass radius, are not well-constrained.  It is of course crucial to ensure that models and observations are comparable.  That is, a model with a long crossing-time might be appropriate for the so-called `leaky' clusters \citep{pfa09}, while it severely underestimates the rate of dynamical evolution of `starburst' clusters such as Wd~1 and the Arches.

The observed $M_{dyn}/m_{lum}$ ratio is subject to its own uncertainties, related to 
{\it (1)} the cluster luminous mass estimate, $M_{lum}$ (see the exemple of Wd~1 in the above discussion),
{\it (2)} the measured velocity dispersion, $\sigma_{los}^2$, and 
{\it (3)} the observed half-light radius.  Here, an additional source of complication is that the half-light radius is stellar-mass dependent if the cluster is primordially mass-segregated \citep[which may be the case for NGC 3603; see ][]{roc10}.  Note that the models presented here are not primordially mass-segregated since the $N$-body simulations of \citet{bau07} build on equal-mass particles.  [Models for the evolution of primordially mass-segregated gas-free clusters are presented in \citet{fle06} and the impact of stellar-evolutionary mass-losses is studied in \citet{ves09}].  

Figure \ref{fig:mdml} clearly demonstrates that star clusters as young as a few Myr observed in virial equilibrium and gas-expulsion driven cluster evolution are not mutually exclusive.  We encourage observers to estimate and quote the crossing-time of the star clusters they study, as well as the size of the field of view they scrutinize.  Both parameters play a pivotal role in assessing reliably the dynamical state of a star cluster.  Finally, we caution that in Figs~\ref{fig:mdml} and \ref{fig:mdlv} the cluster dynamical mass is obtained from the {\it radial} velocity dispersion.  Models of the time-evolution of $M_{dyn}^{pm}/M_{lum}$, where $M_{dyn}^{pm}$ builds on star {\it proper motions}, are highly desirable.   

\section{Future Work and Conclusions}
\label{sec:conclu}

We have studied the impact of cluster-mass-dependent evolutionary rates upon the evolution of the cluster mass function through violent relaxation.  To highlight this so far overlooked process, we have built model star cluster systems with mass-independent star formation efficiency ($SFE$), gas-expulsion time-scale ($\tau_{gexp}/\tau_{cross}$) and tidal field impact ($r_h/r_t$).  Due to the resulting mass-independent infant weight-loss at the end of violent relaxation, the final cluster mass function and the embedded-cluster mass function have the same shape.  Yet, that does not necessarily imply that the cluster mass function {\it during} violent relaxation retains the same shape as that of embedded clusters.  If the evolutionary rate is mass-dependent, it transiently distorts the cluster mass function.

Mass-dependent evolutionary rates arise when the CFRg mass-radius relation is not one of constant mean volume density (since $\tau_{cross} \propto \rho_{CFRg}^{-1/2}$).  For instance, for CFRgs with a constant radius  ($r_{CFRg}$ model, bottom panels of Figs~\ref{fig:MF}-\ref{fig:mfb}), more massive clusters evolve faster and complete their violent relaxation earlier than low-mass clusters.  Conversely, in case of CFRgs of constant mean surface density, dynamical evolution is slower in the high-mass regime ($\Sigma_{CFRg}$ model, top panels of Figs~\ref{fig:MF}-\ref{fig:mfb}).  Therefore, to preserve the shape of the cluster mass function during violent relaxation requires CFRgs of constant mean volume density.  In that case, the bound fraction of stars is mass-independent at {\it any age} during violent relaxation ($\rho_{CFRg}$ model, middle panels of Figs~\ref{fig:MF}-\ref{fig:mfb}).  
We caution, however, that the influence of the CFRg mass-radius relation upon the evolving cluster mass function is weakened when the scatter in cluster initial properties is accounted for (e.g. range in SFE, scatter in radius around a mean mass-radius relation, etc; see Fig.~\ref{fig:MFblur}).

A direct comparison of our models to observations still requires additional work, however.  In our simulations, the mass of a cluster is defined as its bound mass, namely, the mass enclosed within its instantaneous tidal radius.  This may differ from the cluster mass inferred by observers.  Due to their gas-expulsion-driven spatial expansion, star clusters develop untruncated power-law density profiles \citep{els87} at young ages.  That is, unbound stars beyond the tidal radius are not yet spatially dissociated from their `parent' clusters.  Therefore, contrary to older clusters, no truncation in the cluster density profile marks its tidal radius.  The observed mass of those young clusters is thus inferred as the cluster stellar mass brighter than the host galaxy background \citep[see][for an introductory work]{lug12}.  Nevertheless, infant weight-loss beyond the tidal radius and cluster spatial expansion are correlated \citep[see e.g.][]{gey01, bau07}.  At a given age, stronger spatial expansion dims the cluster surface brightness.  Therefore, the cluster bound mass fraction and the cluster mass fraction brighter than the background of stars and gas of the host galaxy are correlated too.  We thus expect the same cluster mass function patterns for cluster masses defined based on a surface brightness limit criterion, although the amplitude of the mass function distortions may differ from those found here.   
In spite of this caveat, it is clear that our work has highlighted, again, the richness of the cluster mass function as a tracer of star cluster formation conditions.

On the scale of individual clusters, violent relaxation can be traced by the evolution of the ratio between the dynamical and luminous masses of a cluster, $M_{dyn}/M_{lum}$.  Models of $M_{dyn}/M_{lum}$ as a function of dynamical time have been computed for various star formation efficiencies and gas-expulsion time-scales (Fig.~\ref{fig:mdml}).  Our simulations range from disrupted clusters to clusters barely affected by gas expulsion.  We insist that for a given set of parameters (SFE, $\tau_{gexp}/\tau_{cross}$, $r_h/r_t$), the rate of dynamical evolution is determined by the CFRg crossing-time. 

Our study has quantified  an effect ignored so far: the impact of the size of the field-of-view on the observed $M_{dyn}/M_{lum}$ ratio.  Smaller apertures are conducive to smaller $M_{dyn}/M_{lum}$ ratios for the following two reasons {\it (1)} Cluster inner regions  return to equilibrium at a time when outer layers still expand through the tidal radius (Fig.~\ref{fig:ap}).  {\it (2)} A smaller aperture `sees' a smaller fraction of the stars expanding beyond the cluster tidal radius (i.e. at large distance from the cluster centre, the only expanding stars to be observed are those in the vicinity of the line-of-sight), which decreases the measured radial velocity dispersion $\sigma^2$.  Both effects lower the mass ratio $M_{dyn}/M_{lum}$, that is,  
a cluster observed after gas expulsion through a smaller aperture is perceived as closer to dynamical equilibrium and its return to equilibrium is seen as faster (Fig.~\ref{fig:mdml}).  We stress that what actually matters is the size of the aperture in units of the cluster initial half-mass radius.  
If the cluster initial crossing-time is short and the observed field-of-view is no larger than a few pc, it is indeed possible to infer $M_{dyn} \simeq M_{lum}$ even though the cluster experienced significant gas expulsion a few Myr earlier.  

Both the cluster mass function and cluster dynamical mass aspects of our work emphasize the crucial importance of the crossing-time to the evolutionary rate of young star clusters.  We strongly encourage observers to estimate and quote the crossing-time of the young star clusters they study, although the observed crossing-time may differ from the initial one by a factor of a few.  From a dynamical point of view, the age of a star cluster expressed in physical-time units hardly tells anything about its evolutionary stage.  The relevant quantity is the age expressed in units of the cluster {\it initial} crossing-time, that is, the CFRg crossing-time (see Fig.~\ref{fig:mrr}).  To obtain a reliable estimate of the latter is thus crucial when it comes to combining  dynamical and photometric modellings of star clusters, as their respective time-units, the initial crossing-time and the year, differ (e.g. see Fig.~\ref{fig:mdlv} for a conversion of the $M_{dyn}/M_{lum}$ ratio into the luminosity-to-dynamical mass ratio).  

The mass-radius relation of CFRgs -- both its slope and normalization -- determines the initial crossing-time of star clusters.  As such, it remains central to our understanding of star cluster dynamical evolution after residual star-forming gas expulsion.

\section{Acknowledgments}
G.P. acknowledges support from the Max-Planck-Institut f\"ur Radioastronomie (Bonn) in the form of a Research Fellowship.  G.P. also thanks Karl Menten, her head of group, for making this research possible.  H.B. acknowledges support from the Australian Research Council through Future Fellowship grant FT0991052.  We thank Susanne Pfalzner and Karl Menten for a careful reading of the initial version of our manuscript.

\end{document}